\documentclass[aps,prl,reprint]{revtex4-1}

\usepackage{graphicx}
\usepackage{dcolumn}
\usepackage{amsmath}
\usepackage{mathrsfs}
\usepackage{amsbsy}
\usepackage{amssymb}
\usepackage{mathtools}
\usepackage{upgreek}

\usepackage[twoside]{rotating}
\usepackage{psfrag}
\usepackage{stackengine}

\usepackage{color,soul}
\definecolor{LightGray}{gray}{0.8}
\definecolor{Orange}{rgb}{1.0, 0.31, 0.0}
\definecolor{Green}{rgb}{0.3, 1.0, 0.3}
\definecolor{Blue}{rgb}{0.75,0.75,1}

\newcommand{\fig}[1]{Fig.~\ref{#1}}
\newcommand{\figs}[1]{Figs.~\ref{#1}}

\newcommand{\bea}{\begin{eqnarray}}
\newcommand{\beal}[1]{\begin{eqnarray}\label{#1}}
\newcommand{\eea}{\end{eqnarray}}
\def\balg#1#2\ealg{\begin{align}\label{#1}#2\end{align}}
\def\balgnl#1\ealgnl{\begin{align*}#1\end{align*}}

\newcommand{\E}{{\mathbf E}}

\newcommand{\B}{{\mathbf B}}

\newcommand{\e}{\boldsymbol{\epsilon}}

\newcommand{\ux}{{\mathbf e}_x}
\newcommand{\uy}{{\mathbf e}_y}
\newcommand{\uz}{{\mathbf e}_z}

\renewcommand{\fig}[1]{Fig.~\ref{#1}}

\renewcommand{\figs}[1]{Figs~\ref{#1}}

\let\originaleqref=\eqref
\renewcommand{\eqref}{Eq.~\originaleqref}

\begin{document}


\title{Three-Dimensional Invisibility to Superscattering Induced by Zeeman-Split Modes}


\author{Grigorios P. Zouros}
\affiliation{Section of Condensed Matter Physics, National and Kapodistrian University of Athens, Panepistimioupolis, GR-157 84 Athens, Greece}
\affiliation{School of Electrical and Computer Engineering, National Technical University of Athens, GR-157 73 Athens, Greece}

\author{Georgios D. Kolezas}
\affiliation{School of Electrical and Computer Engineering, National Technical University of Athens, GR-157 73 Athens, Greece}

\author{Evangelos Almpanis}
\affiliation{Section of Condensed Matter Physics, National and Kapodistrian University of Athens, Panepistimioupolis, GR-157 84 Athens, Greece}
\affiliation{Institute of Nanoscience and Nanotechnology, NCSR ``Demokritos'', Patriarchou Gregoriou and Neapoleos Street, Ag. Paraskevi, GR-153 10 Athens, Greece}

\author{Kosmas L. Tsakmakidis}
\affiliation{Section of Condensed Matter Physics, National and Kapodistrian University of Athens, Panepistimioupolis, GR-157 84 Athens, Greece}



\date{\today}

\begin{abstract}
We report that the fundamental three-dimensional (3-D) scattering single-channel limit can be overcome in magneto-optical assisted systems by inducing nondegenerate magnetoplasmonic modes. In addition, we propose a 3-D active (magnetically assisted) forward-superscattering to invisibility switch, functioning at the same operational wavelength. Our structure is composed of a high-index dielectric core coated by indium antimonide (InSb), a semiconductor whose permittivity tensorial elements may be actively manipulated by an external magnetic bias $\B_0$. In the absence of $\B_0$, InSb exhibits isotropic epsilon-near-zero (ENZ) and plasmonic behavior above and below its plasma frequency, respectively, a frequency band which can be utilized for attaining invisibility using cloaks with permittivity less than that of free space. With realistic $B_0$ magnitudes as high as $0.17$~T, the gyroelectric properties of InSb enable the lift of mode degeneracy, and the induction of a Zeeman-split type dipolar magnetoplasmonic mode that beats the fundamental single-channel limit. This all-in-one design allows for the implementation of functional and highly tunable optical devices.
\end{abstract}


\maketitle


Dynamically manipulating the scattering characteristics of resonant structures comprises a valuable platform for the development of functional, tunable optical devices enabling a variety of applications \cite{kuz_mir_bro_kiv_luk_16,tsa_hes_boy_zha_17}, including optical sorting \cite{shi_lyu_shc_lap_fed_17}, visible-to-invisible switching \cite{ryb_sam_kap_fil_bel_kiv_lim_17}, cloaking-to-superscattering switching \cite{hua_she_min_ver_18}, directionality inversion \cite{arr_mar_pin_16,liu_zha_wan_18,zouros2020magnetic}, all-optical switching \cite{shc_et_al_15}, multipolar interference \cite{lim_ryb_pod_kiv_17}, quantum emitters \cite{cih_cur_raz_kik_bro_18}, or light-matter interaction enhancement at the subwavelength regime via superscattering operation \cite{qia_lin_yan_xio_wan_li_kam_zha_che_19}. A pathway to actively manipulating the scattering response is through external-agent based alteration of the material properties of the structure, which retains the geometry and operational wavelength fixed. Techniques to achieve this dynamic switching between different states include the electrically configurable liquid crystal based meta-optic devices which change the phase or the orientation of the surrounding liquid crystal \cite{kom_et_al_17}, the use of phase-change materials that allow for a permittivity change by transiting between amorphous and crystalline states using external laser beams \cite{lep_kra_alu_19}, magneto-optical media whose permittivity can be modified by an external magnetic bias \cite{kam_ros_pin_far_13}, ferromagnetic materials whose permeability can be manipulated via temperature variations \cite{arr_mar_pin_15}, or ultrafast nonlinear optical switching based on light-induced change of a material's dielectric permittivity (normally, to low values) \cite{cas_et_al_16}.

In this work, we demonstrate that in optimized 3-D core-shell spherical particles consisting of a high-index
dielectric core and an InSb semiconductor coating, we can overcome the fundamental single-channel limit of the scattering efficiency \cite{tri_luk_06}, by appropriately inducing a magnetoplasmonic resonance. In addition, while keeping the geometry and operational wavelength fixed, we can transform the configuration to an invisible state simply by switching off the external magnetic field $\B_0$, utilizing the dynamic properties of the structure's active semiconductor coating, InSb, whose permittivity can be manipulated by relatively small or modest values of $\B_0$. In the presence of low $B_0$ magnitudes, the coating exhibits gyroelectric along with plasmonic activity, resulting in the induction of magnetoplasmonic modes \cite{chen2015,chochol2017} which---different from recent two-dimensional (2-D) superscattering schemes \cite{qia_lin_yan_xio_wan_li_kam_zha_che_19}---are here nondegenerate, i.e., of Zeeman-split type \cite{alm_18,alm_zou_pan_tsa_pap_ste_20,zouros2020magnetic}. We show that these nondegenerate magnetoplasmonic resonances exhibit strong directional scattering. In the absence of $\B_0$, and for the same operational wavelength, InSb exhibits isotropic ENZ and plasmonic behavior above and below its plasma frequency, respectively, a frequency band that can be utilized for attaining invisibility \cite{alu_eng_05,tsa_res_alm_zou_moh_saa_soh_fah_ete_boy_alt_19}. This {\it off}-to-{\it on} transition from invisibility to superscattering operation is robust under material losses and gives rise to a huge enhancement in the scattering efficiency. The findings obtained from our formal analytical solution are in excellent agreement with full-wave simulations (COMSOL), thereby establishing the validity of the proposed set-up.

\begin{figure}[!t]
\centering
\includegraphics[scale=1]{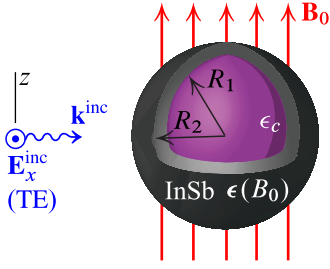}
\vspace{-4mm}
\caption{Schematic representation of the proposed set-up. Linear TE polarized light impinges at an angle of $90^\circ$ with respect to the external magnetic field ${\mathbf B}_0$.}
\label{setup}
\end{figure}

In \fig{setup} we show the configuration of the forward-superscattering to invisibility
magnetic switch. The main structure consists of a $R_1$-radius high permittivity $\epsilon_c=20\epsilon_0$ dielectric core---with $\epsilon_0$ being the free space permittivity---coated by a $R_2$-radius shell consisting of semiconductor InSb. The whole configuration is exposed in external magnetic field $\B_0=B_0\uz$. Since $\B_0$ is $z$-oriented, InSb's permittivity tensor is expressed in cartesian coordinates by $\e(B_0)=\epsilon_1(B_0)(\ux\ux^T+\uy\uy^T)+i\epsilon_2(B_0)(\ux\uy^T-\uy\ux^T)+\epsilon_3(B_0)\uz\uz^T$---where $T$ denotes transposition---with $\epsilon_1(B_0)=\epsilon_0\epsilon_\infty\{1-(\omega+iv)\omega_p^2/\omega/[(\omega+iv)^2-\omega_c^2]\}$, $\epsilon_2(B_0)=\epsilon_0\epsilon_\infty\{\omega_c\omega_p^2/\omega/[(\omega+iv)^2-\omega_c^2]\}$ and $\epsilon_3(B_0)=\epsilon_0\epsilon_\infty[1-\omega_p^2/\omega/(\omega+iv)]$ \cite{tsa_she_sch_zhe_uph_den_alt_vak_boy_17}. In the aforementioned relations we have used realistic material parameters where $\epsilon_\infty=15.6$ accounts for interband transitions, $\omega_p=(N_ee^2/\epsilon_0/\epsilon_\infty/m^\ast)^{1/2}=4\pi\times10^{12}$~rad/s is the plasma angular frequency (with $N_e$ the electron density, $e$ the elementary charge and $m^\ast=0.0142m_e$ electron's effective mass, where $m_e$ is electron's rest mass), $\omega_c=eB_0/m^\ast$ is cyclotron angular frequency, and $v=0.001\omega_p$ the damping angular frequency which accounts for losses. In case of null $\B_0$, InSb turns isotropic with $\epsilon_1(0)\equiv\epsilon_3(0)$ and $\epsilon_2(0)=0$. In addition, the core-shell structure is nonmagnetic and it is located in free space. The set-up is illuminated by a normal to $\B_0$ impinging transverse electric (TE) plane wave---i.e., the incident electric field $\E^{\rm inc}$ is normal to the plane of incidence, therefore, if the plane of incidence is the $yz$-plane as shown in \fig{setup}, then $\E^{\rm inc}$ is $x$-polarized. To examine the electromagnetic (EM) response of the magnetic switch proposed in this work, the solution of EM plane wave scattering by dielectric-gyroelectric core-shell spheres is necessary. This is feasible by the discrete eigenfunction method employed in \cite{li_ong_11,kol_zou_tsa_19} for homogeneous spheres and spheroids. In the Supplemental Material \cite{supplemental} we extend that method and we rigorously develop a full-wave analytical solution of EM scattering by dielectric-gyroelectric spheres. The scattering efficiency is given by
\balg{1}
Q_{sc}=\frac{\lambda_0^2}{\pi}\sum_{m=-n}^n\sum_{n=1}^\infty \Big( |a_{mn}|^2 + |b_{mn}|^2 \Big),
\ealg
where $\lambda_0$ is the free space wavelength, $a_{mn}$, $b_{mn}$ the expansion coefficients of the scattered electric field \cite[Eq.~(S2)]{supplemental}, and $m$, $n$ the spherical harmonics and angular momentum indices, respectively. In the long wavelength limit, the spectrum is dominated by the electric dipolar (ED) and magnetic dipolar (MD) responses for which $n=1$. In particular, for purely isotropic scatterers, the ED response to the scattering efficiency is obtained when only the $b_{11}$ expansion coefficient contributes in \eqref{1}, while the MD response stems exclusively from $a_{11}$.

When the MD response is negligible and the ED response dominates the spectrum, transparency can be achieved by utilizing an isotropic lossless ENZ coating with permittivity $\epsilon_1<\epsilon_0$ and a radii ratio determined by \cite{alu_eng_05}
\balg{2}
\frac{R_1}{R_2}=\Big[\frac{(\epsilon_1-\epsilon_0)(2\epsilon_1+\epsilon_c)}{(\epsilon_1-\epsilon_c)(2\epsilon_1+\epsilon_0)}\Big]^{1/3}.
\ealg
By appropriately selecting $\epsilon_1$, $\epsilon_c$, \eqref{2} yields the radii ratio for which cancellation of the ED term to the spectrum is achieved. \eqref{2} is ideally suited for lossless and non-dispersive materials. 
To make a rough estimation of a possible transparency window,
using \eqref{2}, we neglect losses by setting zero damping term $v$ in Drude-Lorentz model of InSb, we set $\epsilon_1=0.4929\epsilon_0$ as an average value of $\epsilon_1$ in the range from $2$~THz to $2.065$~THz, and along with $\epsilon_c=20\epsilon_0$, \eqref{2} yields $R_1/R_2=0.650$. 
This radii ratio can serve as a tentative basis also in our case, where the InSb coating is lossy and dispersive. 

\begin{figure}[!t]
\centering
\includegraphics[scale=0.91]{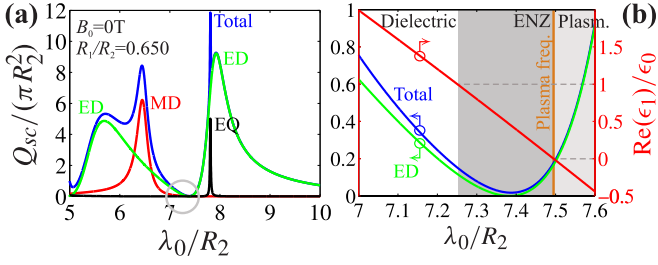}
\vspace{-4mm}
\caption{Achieving ENZ- and plasmonic-based transparency under null $\B_0$ in the set-up of \fig{setup} with $\epsilon_c=20\epsilon_0$, $R_2=20~\mu$m and $R_1/R_2=0.650$.
(a) Spectrum in the subwavelength regime. Blue: total normalized $Q_{sc}$; red: MD term; green: ED term; black: EQ term. The gray circle indicates the region where transparency is achieved.
(b) Zoom in the $\lambda_0/R_2$ window indicated by the gray circle in (a). Blue/left axis: total normalized $Q_{sc}$; green/left axis: ED term; red/right axis: ${\rm Re}(\epsilon_1)/\epsilon_0$ of the InSb coating above and below plasma frequency $f_p=2$~THz (shown by the vertical orange line at $\lambda_0/R_2=7.495$). White region (dielectric): ${\rm Re}(\epsilon_1)/\epsilon_0>1$; dark-gray region (ENZ): $0<{\rm Re}(\epsilon_1)/\epsilon_0<1$; bright-gray region (plasmonic): ${\rm Re}(\epsilon_1)/\epsilon_0<0$.
}
\label{spec}
\end{figure}

At first, we examine the optical response of the lossy and dispersive core-shell particle in the absence of $\B_0$, where InSb is isotropic. 
By employing the radii ratio $R_1/R_2=0.650$, we obtain the spectrum shown in \fig{spec}(a).
In blue line we show the total normalized $Q_{sc}$ which involves all terms of \eqref{1}, up to an index $n_{\rm max}$  for which convergence is ensured.
In addition, we also depict the separate MD/ED and electric quadrupolar (EQ) terms. The latter one contributes to \eqref{1} by employing only the $a_{12}$ expansion coefficient. The ED/MD resonances appearing in the range from $\lambda_0/R_2=5$ up to $\lambda_0/R_2=7$ in \fig{spec}(a), are typical subwavelength resonances due to the high-index dielectric core. On the contrary, the EQ/ED peaks above $\lambda_0/R_2=7.5$ are plasmonic resonances of the coating. This is because at frequencies below the plasma frequency $f_p=\omega_p/(2\pi)$, InSb changes its properties from ENZ to plasmonic. %
This is illustrated in \fig{spec}(b) which focuses in the transition region indicated by the gray circle in \fig{spec}(a). There, the two domains (ENZ/plasmonic) are marked by dark-gray and bright-gray colors, the boundary of which is determined by the plasma frequency at $\lambda_0/R_2=7.495$ (or $f_p=2$~THz). 
The real part of the coating permittivity ${\rm Re}(\epsilon_1)$ depends almost linearly on the wavelength, as shown in \fig{spec}(b). In particular, ${\rm Re}(\epsilon_1)/\epsilon_0=0$ at $\lambda_0/R_2=7.495$ (which corresponds to frequency $f_0$ equal to $f_p$), while ${\rm Re}(\epsilon_1)/\epsilon_0\approx1$ at $\lambda_0/R_2=7.25$ ($f_0=2.068$~THz). 
In brief, \fig{spec}(b) indicates that inside the ENZ region of the coating, the normalized $Q_{sc}$ from solely the ED contribution (which is similar to the total normalized $Q_{sc}$) is suppressed, resulting in a respective transparency window. However, as we shall see below, this is not the case when $\B_0 \ne 0$.

\begin{figure}[!t]
\centering
\includegraphics[scale=0.91]{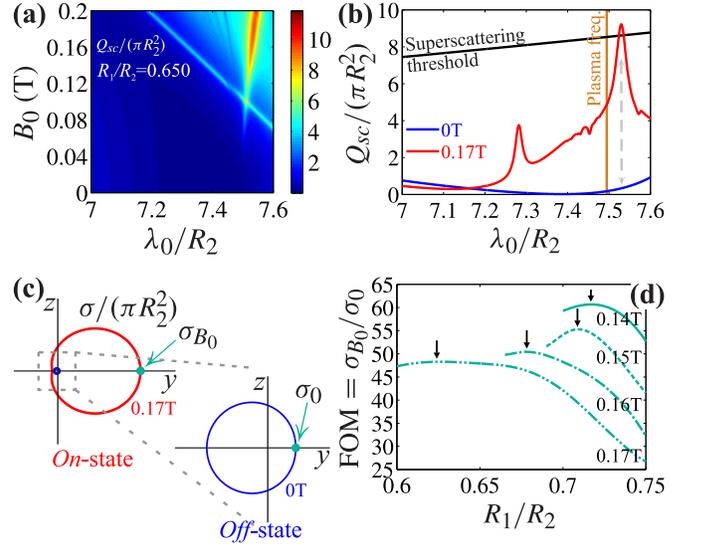}
\vspace{-4mm}
\caption{Application of external $\B_0$ and 2-D optimization. The set-up is the same as in \fig{spec}.
(a) Total normalized $Q_{sc}$ vs $B_0$ and $\lambda_0/R_2$ in the transparency-enabled $\lambda_0/R_2$ window as indicated by \fig{spec}(b).
(b) Comparison of total normalized $Q_{sc}$ between null and non-null external $\B_0$ states. Blue: $B_0=0$~T; red: $B_0=0.17$~T; black: superscattering threshold. The gray dashed arrowhead line depicts the precise $\lambda_0/R_2=7.530$ of the {\it off}-{\it on} transition.
(c) Normalized $\sigma$ on $yz$-plane when the structure operates at $\lambda_0/R_2=7.530$. Red: {\it on}-state/forward-superscattering for $B_0=0.17$~T; blue: {\it off}-state/transparency for $B_0=0$~T.
(d) 2-D optimization of FOM vs $B_0$ and $R_1/R_2$. The small arrows indicate the optimal states at which FOM is maximized.}
\label{map}
\end{figure}

In \fig{map}(a) we show the map of the normalized $Q_{sc}$, versus the external $B_0$ magnitude, in the same wavelength window as the one in \fig{spec}(b). All the other parameters of the configuration are the same with the ones used in \fig{spec}.
The deep blue region at $B_0=0$~T
in \fig{map}(a) corresponds to the transparency window discussed in \fig{spec}(b), where normalized $Q_{sc}$ is almost zero.
By {\it switching on} the external magnetic field, and varying its magnitude, we can monitor the impact that it has on this transparency window.
In particular, for $B_0 \ne 0$~T, sharp peaks in intense red color appear, which correspond to high $Q_{sc}$ values. 
In \fig{map}(b) we plot the normalized $Q_{sc}$ for two particular choices of $B_0$, i.e., $B_0=0$ and $B_0=0.17$~T. The first case ($B_0=0$)
is already discussed but shown for comparison. 
At $B_0=0.17$~T, two dominant scattering modes appear, where the longer-wavelength one has greater normalized $Q_{sc}$ than the dipole-superscattering single-channel threshold given by \cite{tri_luk_06}
\balg{3}
\frac{Q_{sc}}{\pi R_2^2}=\frac{3}{2\pi^2}\Big(\frac{\lambda_0}{R_2}\Big)^2.
\ealg
The corresponding dipole-superscattering threshold is shown in \fig{map}(b) in black. 
As shown, the superscattering resonance at $\lambda_0/R_2=7.530$ 
can be dramatically suppressed by {\it switching off} the external magnetic field, where the particle becomes almost transparent. This transition from superscattering operation---$Q_{sc}/(\pi R_2^2)=9.223$ ($B_0=0.17$~T) to almost transparency---$Q_{sc}/(\pi R_2^2)=0.3339$ ($B_0=0$~T), controlled entirely by the external magnetic bias, is pointed out with a dashed arrow in \fig{map}(b).       
In \fig{map}(c) we depict the radiation pattern of the structure when the latter operates exactly at $\lambda_0/R_2=7.530$, by plotting the normalized bistatic scattering cross section $\sigma(\theta,\varphi)/(\pi R_2^2)$ on $yz$-plane by varying $\theta\in[0,\pi]$, for the two different $B_0$ values that allow for the {\it on}-{\it off} switching. When $B_0=0.17$~T, the coated particle operates at a forward-superscattering state with $\sigma_{B_0}/(\pi R_2^2)\equiv\sigma(\theta=\pi/2,\varphi=\pi/2)/(\pi R_2^2)=20.45$, i.e., the value of $\sigma/(\pi R_2^2)$ in the forward direction. Contrariwise, when $B_0=0$~T, the particle is at {\it off}-state/transparency with the respective point in the suppressed radiation pattern being $\sigma_0/(\pi R_2^2)\equiv\sigma(\theta=\pi/2,\varphi=\pi/2)/(\pi R_2^2)=0.4269$. 
We point out here that the lower-wavelength and smaller in amplitude resonance in \fig{map}(b) at $\lambda_0/R_2=7.282$, when $B_0=0.17$~T,
is much below the single-channel limit. Additionally, this mode does not exhibit unidirectional scattering. In particular, this resonance shows a poor $\sigma_{B_0}/(\pi R_2^2)=5.48$ while, at the same instance, it has non negligible backscattering response $\sigma(\theta=\pi/2,\varphi=3\pi/2)/(\pi R_2^2)=2.08$ rendering it improper for the application studied here. 


Next, we define a figure of merit (FOM) as the ratio
\balg{4}
{\rm FOM}=\frac{\sigma_{B_0}}{\sigma_0}.
\ealg
For the specific example of \fig{map}(c),
FOM is not optimized, and
yields ${\rm FOM}=47.90$. In \fig{map}(d) we show that for a given $B_0$ there exists an optimal $R_1/R_2$ ratio for which FOM is maximized, yet the superscattering state is maintained. This is regarded as a 2-D optimization of FOM vs $B_0$ and $R_1/R_2$. \fig{map}(d) shows four constant-$B_0$ curves of FOM vs $R_1/R_2$ for $0.14$~T, $0.15$~T, $0.16$~T and $0.17$~T. All values in \fig{map}(d) correspond to greater normalized $Q_{sc}$ than the fundamental single-channel limit. For the preceding example of $B_0=0.17$~T, optimal FOM is achieved at $R_1/R_2=0.625$. The other maximum values shown by the arrows in \fig{map}(d) correspond to $R_1/R_2=0.675$ ($B_0=0.16$~T), $R_1/R_2=0.705$ ($B_0=0.15$~T) and to $R_1/R_2=0.715$ ($B_0=0.14$~T).

\begin{figure}[!t]
\centering
\includegraphics[scale=0.91]{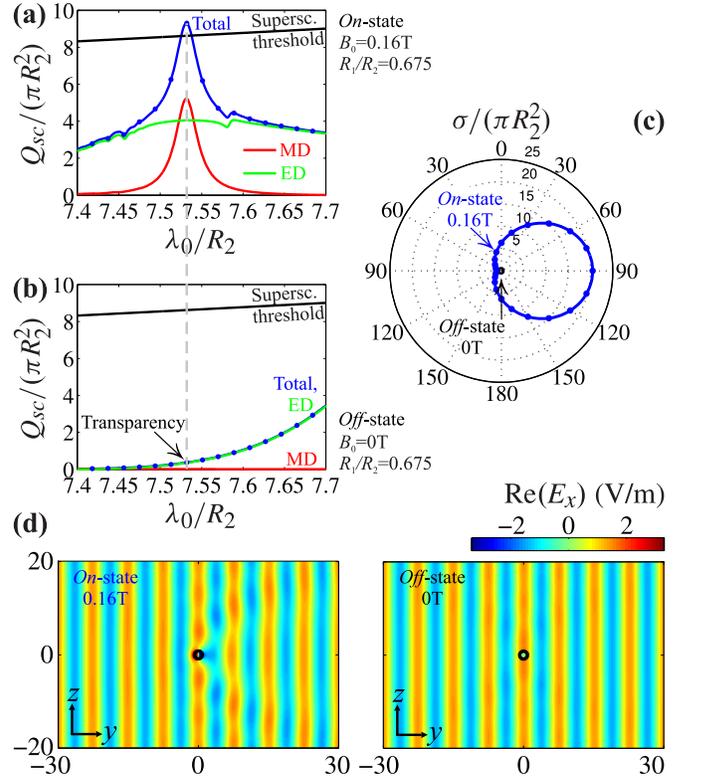}
\vspace{-4mm}
\caption{Engineering the magnetic switch for optimal operation at $R_1/R_2=0.675$ and $B_0=0.16$~T. The other values of parameters are the same as in \fig{spec}.
(a) Spectrum in {\it on}-state/superscattering operation with $B_0=0.16$~T. Blue: total normalized $Q_{sc}$/curve: our method/dots: COMSOL; red: MD term due to $a_{0,1}$; green: ED term due to $b_{-1,1}$; black: superscattering threshold.
(b) Spectrum in {\it off}-state/transparency operation with $B_0=0$~T. Blue: total normalized $Q_{sc}$/curve: our method/dots: COMSOL; red: MD term; green: ED term; black: superscattering threshold.. The gray dashed vertical line depicts the value $\lambda_0/R_2=7.532$ where the optimal {\it off}-{\it on} transition takes place.
(c) Radiation pattern of normalized $\sigma$ on $yz$-plane under $\lambda_0/R_2=7.532$ excitation. Blue: {\it on}-state/forward superscattering when $B_0=0.16$~T/curve: our method/dots: COMSOL; black: {\it off}-state/transparency when $B_0=0$~T.
(d) Near-field of ${\rm Re}(E_x)$ on $yz$-plane under {\it on}-state (left) and {\it off}-state (right) activity.}
\label{opt}
\end{figure}


In \fig{opt} we demonstrate the realization of the invisibility to superscattering magnetic switch for the optimized structure ($R_1/R_2=0.675$) at $B_0=0.16$~T. \fig{opt}(a) depicts the total normalized $Q_{sc}$ for the {\it on}-state operation, along with its dipolar components. Higher-order terms---such as quadrupolar---are negligible in this regime. We note in passing that,
for an isotropic spherical particle, 
the $m$-mode index in \eqref{1}
is degenerate, which means that a multipolar mode of electric or magnetic type is only characterized by the angular momentum index $n$.  
However, this is not the case
for gyroelectric spherical particles, as in the present case, where the lifting of the $m$-degeneracy leads to $2n+1$ separate Zeeman-split modes \cite{alm_18,alm_zou_pan_tsa_pap_ste_20}.
Since in the $\lambda_0/R_2$ regime of \fig{opt}(a) only the dipolar  ($n=1$) modes exist, we
expect three ($2n+1=3$) nondegenerate modes for each dipolar (parent) mode, with indices $m=0,\pm1$.
However, the coupling of each one of such modes with the external radiation strongly depends on the angle of incidence and polarization~\cite{var_ste_16,alm_18}.  
Our full-wave analytical calculations reveal that, in the specific spectral window of \fig{opt}(a), for light impinging as shown in \fig{setup}, we have
$a_{\pm1,1}=b_{0,1}=b_{1,1}=0$ while the nonzero expansion coefficients are only $a_{0,1}$ and $b_{-1,1}$ and yield nondegenerate MD and ED modes, respectively.
The contribution of $a_{0,1}$ and $b_{-1,1}$ to the normalized $Q_{sc}$ is depicted in \fig{opt}(a) by the red and green curves, respectively. As evident, 
the high peak at $\lambda_0/R_2=7.532$ beats the superscattering limit. The physical mechanism behind this scattering maximization is attributed to the overlapping---as shown by the dashed gray line---of the nondegenerate $m=0$ MD and $m=-1$ ED contributions, and is understood
as a magnetoplasmon polariton excited in the magnetized InSb coating~\cite{bri_wal_har_bur_72,xia_zha_wan_zha_han_19}. Obviously, this is not the case when InSb is not magnetized, as in the case of \fig{spec}(a) where the coating exhibits purely plasmonic properties and the specrtum is dominated by degenerate ED/EQ resonaces only. We note in passing that, the plasmonic Fano feature~\cite{luk_zhe_mai_hal_nor_gie_cho_10} at $\lambda_0/R_2=7.587$---as shown in  \fig{opt}(a)---is attributed to the nondegenerate $m=-1$ ED contribution, where a similar asymmetric feature has been also observed in ring-type cavities \cite{hao_son_dor_mai_hal_nor_08}. The respective spectrum at {\it off}-state 
is shown in \fig{opt}(b),
using the same scale as the one in \fig{opt}(a). At $\lambda_0/R_2=7.532$, the system is almost transparent with normalized $Q_{sc}$ being significantly suppressed at $Q_{sc}/(\pi R_2^2)=0.3746$. With a $Q_{sc}/(\pi R_2^2)=9.302$ when $B_0=0.16$~T, this {\it off}-to-{\it on} transition results to a $2383\%$ enhancement in $Q_{sc}$. \fig{opt}(c) depicts the radiation pattern when the system operates at $\lambda_0/R_2=7.532$. The forward-superscattering operation is achieved with a ${\rm FOM}=50.38$ and an almost null radiation pattern when $B_0=0$~T. To validate our findings, we fully compare our analytical solution with COMSOL's finite-element solver. The total normalized $Q_{sc}$, as obtained by our method, is in full agreement with COMSOL, with the latter being depicted by the blue dots in \figs{opt}(a)--(c). To have a degree of comparison, our method yields $Q_{sc}/(\pi R_2^2)=9.302$ and $0.3746$ at the magnetoplasmonic resonance and at the transparency state, respectively, when $\lambda_0/R_2=7.532$. COMSOL's respective values are $Q_{sc}/(\pi R_2^2)=8.892$ and $0.3746$ while the superscattering threshold is $Q_{sc}/(\pi R_2^2)=8.622$. In addition, the radiation pattern depicted in \fig{opt}(c) is in full agreement with COMSOL. \fig{opt}(d) depicts the ${\rm Re}(E_x)$ on $yz$-plane for {\it on}- and {\it off}-state activity. \fig{opt}(d)/right confirms the invisibility state since the incoming wave passes through the scatterer almost unperturbed, while the snapshot in \fig{opt}(d)/left clearly illustrates the forward-scattering propagation along the positive $y$-axis.

In conclusion, based on a rigorous analytical solution of the 3-D EM scattering problem by dielectric-gyroelectric spheres, corroborated by full-wave simulations, we introduced a high-permittivity/semiconductor structure for breaking the fundamental single-channel limit of the scattering efficiency. This operation is established by inducing nondegenerate dipolar magnetoplasmonic modes, in the subwavelength regime, by operating the structure below the semiconductor's plasma frequency, and by applying a low external magnetic bias in the range $0.14$~T--$0.17$~T. Likewise, for the same operational wavelength and core-shell radii, for which superscattering operation occurs, we have also shown how 3-D invisibility can be attained, simply by turning off the external magnetic field. The latter effect stems from the dynamic properties of the coating, which change to a plasmonic behaviour in the absence of magnetic field, enabling a plasmonic based cloak. By a 2-D optimization relative to the external magnetic field and the radii ratio, we identified optimal states for which the {\it on}-{\it off} ratio of the forward scattering cross sections is maximized, with a concurrent enormous enhancement in the total scattering efficiency. Our findings could pave the way towards the design of highly functional and tunable optical devices, including state-of-the-art optical metasurfaces \cite{nes_aha_18}.

G.P.Z., E.A., and K.L.T. were supported by the General Secretariat for Research and Technology (GSRT) and the Hellenic Foundation for Research and Innovation (HFRI) under Grant No. 1819.

\end{document}